\documentclass{article}
\usepackage[dvips]{graphicx}
\makeatletter
\def\@biblabel#1{#1.}
\makeatother

\begin{document}

{\Large\bf
Anomalous charge transport of the parent antiferromagnet  
Bi$_2$Sr$_2R$Cu$_2$O$_8$}

\vspace*{4mm}
 
I. Terasaki, T. Takemura, T. Takayanagi and T. Kitajima

Department of Applied Physics, Waseda University,
Tokyo 169-8555, JAPAN

\vspace*{5mm}
%--------------------------------------------------
%     Abstract
%--------------------------------------------------

{\bf Abstract:} 
The resistivity and the thermopower for
Bi$_2$Sr$_2$Ca$_{1-x}R_x$Cu$_2$O$_8$ single crystals
are measured and analyzed
with a special interest in the parent antiferromagnet insulators.
Above room temperature, the parent insulators
show an electric conduction confined in the CuO$_2$ plane 
and a decreasing thermopower with temperature, 
which are very similar to those for high-$T_c$ cuprates. 
These data strongly suggest that essential anomalies 
of high-$T_c$ cuprates inheres in the parent insulators.
\vspace*{6mm}

%--------------------------------------------------
%     Introduction
%--------------------------------------------------
{\large\bf INTRODUCTION}

The anisotropic charge transport of high-$T_c$ cuprates (HTSC)
has been one of the biggest anomalies 
that cannot be explained by conventional solid-state theories.
In this context the parent antiferromagnetic (AF) insulators 
are to be carefully investigated as a limit of underdoping.
Since they exhibit a characteristic
two-dimensional (2D) AF fluctuation above the N\'eel temperature $T_N$,
their transport properties are expected to
show a qualitative change below and above $T_N$. 
There appears increasing evidence of 
close resemblance between the AF order and the $d$-wave
superconductivity (SC) \cite{Laughlin,Ronning},
which again implies the anomalous properties of the parent insulators.

In spite of the above prospects, the transport properties of 
the parent insulators have attracted less interest. 
We have been studying the charge transport
of the parent insulator Bi$_2$Sr$_2R$Cu$_2$O$_8$ 
($R$=Y and rare-earth) \cite{Kitajima}
and here we report on the recent results of
the resistivity and the thermopower from 4 to 500 K.

\begin{figure}
\begin{center}
 \includegraphics[width=15cm,clip]{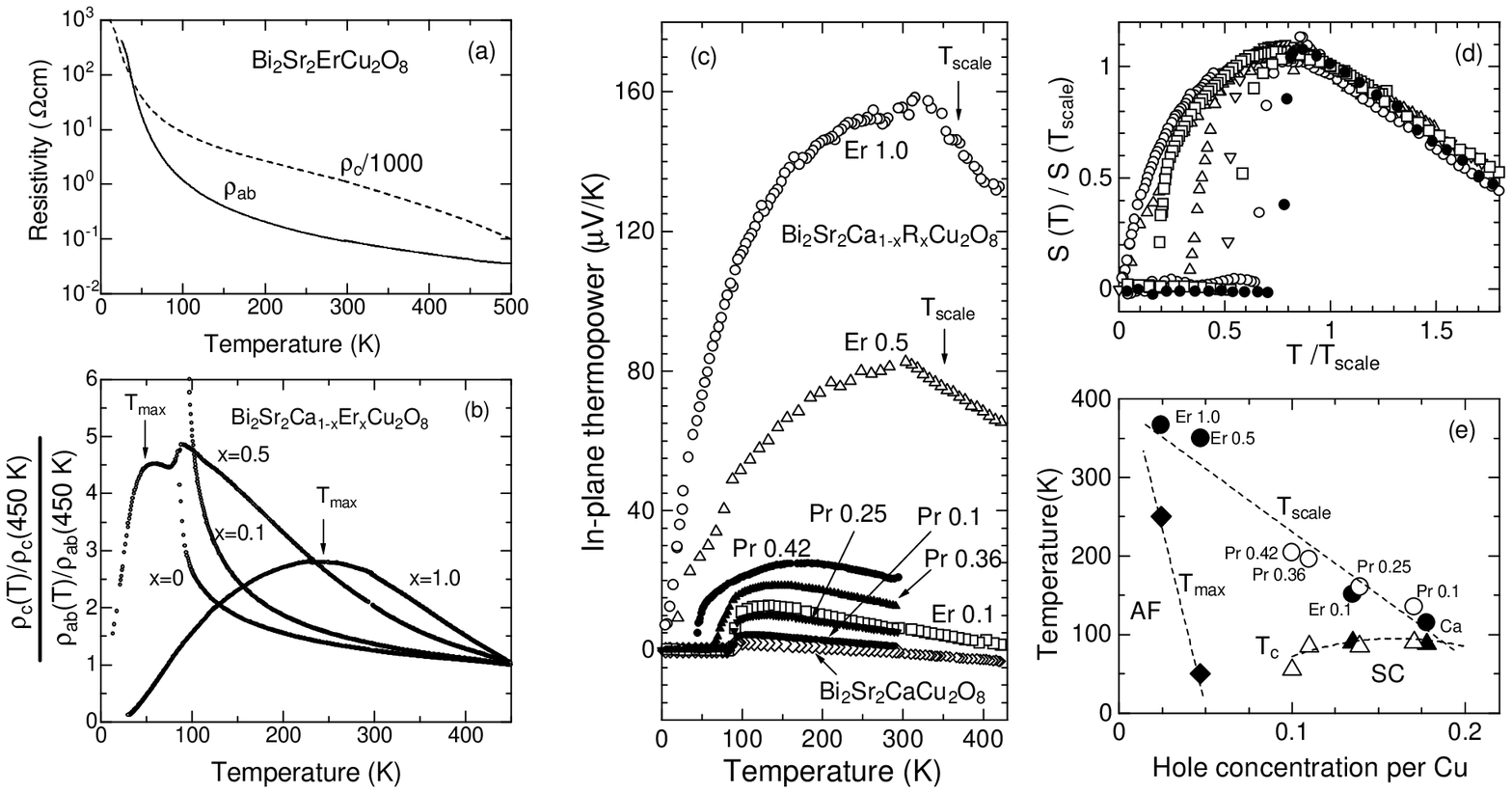}
\end{center}
{\small Fig. 1 \quad 
(a) The in-plane resistivity ($\rho_{ab}$) and 
the out-of-plane  resistivity ($\rho_c$) 
of Bi$_2$Sr$_2$ErCu$_2$O$_8$ single crystals,
(b) $\rho_c/\rho_{ab}$
for Bi$_2$Sr$_2$Ca$_{1-x}$Er$_x$Cu$_2$O$_8$
normalized at 450 K,
(c) The in-plane thermopower of
Bi$_2$Sr$_2$Ca$_{1-x}R_x$Cu$_2$O$_8$,
(d) The same data in Fig. 1(c) normalized at $T_{\rm scale}$, and
(e) $T_{\rm max}$, $T_{\rm scale}$ and $T_c$ plotted as a function of
the hole concentration per Cu estimated from the room-temperature
thermopower.}   
\end{figure}

%--------------------------------------------------
%     Experimental
%--------------------------------------------------

{\large\bf EXPERIMENTAL}

Single crystals of Bi$_2$Sr$_2$Ca$_{1-x}R_x$Cu$_2$O$_8$ were grown
by a flux technique.
The growth conditions and the characterization of the samples
were described elsewhere \cite{Kitajima}.
The resistivity was measured from 4.2 to 500 K
through a four-probe method.
The thermopower was measured 
using a steady-state technique
from 4.2 to 450 K with a temperature gradient of 0.5-1.0 K.
A thermopower of voltage leads was carefully subtracted.

%---------------------------------------------------------
%     Results and Discussion
%---------------------------------------------------------
{\large\bf RESULTS AND DISCUSSION}

Figure 1(a) shows the in-plane resistivity $\rho_{ab}$ 
and the out-of-plane  resistivity $\rho_c$
of Bi$_2$Sr$_2$ErCu$_2$O$_8$ single crystals.
Although $\rho_{ab}$ and $\rho_c$ are insulating, their temperature
dependence is significantly different,
which is clearly seen in the resistivity ratio $\rho_c/\rho_{ab}$.
As shown in Fig. 1(b), $\rho_c/\rho_{ab}$ for
Bi$_2$Sr$_2$Ca$_{1-x}$Er$_x$Cu$_2$O$_8$ systematically evolves with $x$
\cite{Kitajima}.
We should emphasize that $\rho_c/\rho_{ab}$ smoothly changes with $x$ 
above room temperature.
If one looked at $\rho_c/\rho_{ab}$ only above room temperature,
one could not distinguish the parent insulators ($x$=0.5 and 1)
from the superconductors ($x$=0 and 0.1). 
Thus we may say that the holes are
confined in the parent insulator as well as in HTSC.
Furthermore, $\rho_c/\rho_{ab}$ for $x$=1.0 and 0.5
takes a broad maximum at a certain temperature $T_{\rm max}$,
as indicated by arrows in Fig. 1(a).
By comparing the $\mu SR$ data for Bi$_2$Sr$_2$Ca$_{1-x}$Y$_x$Cu$_2$O$_8$
\cite{Nishida},
we find that $T_{\rm max}$ is very close to $T_N$.
These data suggest that the 2D AF fluctuation above $T_N$ 
causes the confinement in the CuO$_2$ plane,
and that the 3D AF order below $T_N$ releases the holes from
the confinement.
In other words, the confinement is effective only in a spin liquid. 
Very recently, Hanke et al. \cite {Hanke} have pointed out that
the AF order in the parent insulator can be regarded as
a Bose-Einstein condensation (BEC) of $S=1$ magnons in the 
resonating-valence-bond background.
In this context, the charge transport for the parent insulator
may probe some anomalies associated with BEC of magnons.

The in-plane thermopower $S$ of the parent insulators 
is also anomalous.
In Fig. 1(c), the thermopowers for 
Bi$_2$Sr$_2$Ca$_{1-x}$Er$_x$Cu$_2$O$_8$
and Bi$_2$Sr$_2$Ca$_{1-x}$Pr$_x$Cu$_2$O$_8$
are plotted as a function of temperature.
$S$ at room temperature monotonically increases with $x$,
which is consistent with the literature \cite{Tallon}.
This indicates that doping levels are 
controlled by the substitution of Er or Pr for Ca.
A significant anomaly is that
$S$ for the parent insulators is roughly expressed by $A-BT$ 
above a temperature $T_{\rm scale}$,
where $A$ and $B$ are positive constants.
Note that $S$ for HTSC also obeys the relation of $A-BT$, while
the diffusive part of the thermopower of conventional conductors will
be independent of temperature at high temperatures for $E_F \ll k_BT$.
Thus $B$ and $T_{\rm scale}$ are another hallmark
that discriminates the parent insulator and HTSC from the other solids.
In addition,  $B$ and $T_{\rm scale}$ is larger for larger $x$,
implying a certain relationship between them.
To see the relationship more clearly we plot
$S(T)/S(T_{\rm scale})$ as a function of $T/T_{\rm scale}$
in Fig. 1(d).
Most unexpectedly, all the curves of $S(T)/S(T_{\rm scale})$ 
fall on a single curve, which clearly indicates
a good scaling relation of the thermopower.

Let us compare $T_{\rm scale}$ with other experiments.
In Fig. 1(e), $T_c$, $T_{\rm max}$ and $T_{\rm scale}$ are plotted as a
function of hole concentration per Cu ($p$) that is estimated through
an empirical relation to the room-temperature thermopower \cite{Tallon}.
$T_{\rm scale}$ linearly decreases with $p$,
which reminds us of the pseudo-gap temperature $T^*$.
In fact, we have found $T_{\rm scale}$ for our superconducting samples 
to be close to $T^*$ observed 
in the STM/STS or photoemission experiments \cite{Takemura}.
Although the physical meaning of $T^*$ is still controversial, 
many physical parameters follow a similar scaling relation.
In particular, the Hall coefficient \cite{Hwang}
and the susceptibility \cite{Nakano}
can be scaled in terms of $T/T^*$,
which strongly suggests that the carrier density or
the density of states is a function of $T/T^*$.
Accordingly the thermopower may also be expressed as a function of $T/T^*$
in the sense that it gives a measure of the carrier density
or the density of states.

The scaling behavior is consistent with a recent photoemission study on 
a parent insulator Ca$_2$CuO$_2$Cl$_2$ \cite{Laughlin,Ronning},
where a $d$-wave-like charge gap is opened in the $k$ space.
This means that the holes are doped initially near
the node ($\pi/2$, $\pi/2$) at low temperatures, 
and the doped region would be somewhat spread around the node 
at high temperature, because the electron correlation
weakened by thermal excitation decreases or smears the charge gap.
This situation indeed resembles 
what is observed in the underdoped HTSC.
At $T$=0, only the node has the zero-energy excitation,
and above $T^*$ the Fermi surface grows from ($\pi/2$, $\pi/2$)
to ($\pi/2$, 0) and (0, $\pi/2$).

As mentioned above, we can capture a gross feature of the phase 
diagram of Bi$_2$Sr$_2$Ca$_{1-x}R_x$Cu$_2$O$_8$ 
by measuring $\rho_{ab}$, $\rho_c$ and $S$,
which give $T_c$, $T_{\rm max}$ ($\sim T_N$) 
and $T_{\rm scale}$ ($\sim T^*$) as a function of $p$.
Hence we would like to emphasize that the resistivity 
and the thermopower are a simple and powerful tool
to explore the phase diagram.
We further claim that
the high-temperature transport is essentially the same between
the parent insulator and HTSC.
Above all, $T_{\rm scale}$ is indicative of the existence 
of the pseudo-gap in the parent insulator,
which should be further examined by different probes.

%---------------------------------------------------------
%    Summary
%---------------------------------------------------------
{\large\bf SUMMARY}

We prepared a set of single crystals of Bi$_2$Sr$_2$Ca$_{1-x}R_x$Cu$_2$O$_8$,
and measured the resistivity and the thermopower.
We have found that the resistivities of the parent insulators 
exhibit a confinement behavior above $T_{\rm max}$ 
(near the N\'eel temperature)
and the thermopower shows a good scaling behavior
for a certain temperature $T_{\rm scale}$ 
(near the pseudo-gap temperature). 
These results strongly suggest a close similarity between
the antiferromagnetic insulators and the high-temperature 
superconductors.

%---------------------------------------------------------
%   Acknowledgments
%---------------------------------------------------------
{\bf Acknowledgments.} 
This work was partially supported
by KAWASAKI STEEL 21st Century Foundation.
The authors would like to thank T. Itoh, T. Kawata, K. Takahata,
Y. Iguchi and T. Sugaya for collaboration.

%---------------------------------------------------------
%    References
%---------------------------------------------------------

\end{document}